\documentstyle[aps,epsf,amssymb]{revtex}

\begin{document}
\draft

\title{
\rightline{\small{\em To appear in Journal of 
	Mathematical Physics (August 2000)\/}}
	Ill-posedness in the Einstein equations
       }

\author{Simonetta Frittelli\thanks{e-mail: 
	simo@mayu.physics.duq.edu}$^{a,b}$
	\and Roberto G\'{o}mez$^b$
}

\address{
	 $^a$Department of Physics, Duquesne University, Pittsburgh, 
		PA 15282\\
	 $^b$ Department of Physics and Astronomy, 
	University of Pittsburgh, Pittsburgh, PA 15260	
	}

\date{\today}

\maketitle

\begin{abstract}

It is shown that the formulation of the Einstein equations widely in
use in numerical relativity, namely, the standard ADM form, as well
as some of its variations (including the most recent
conformally-decomposed version), suffers from a certain but standard
type of ill-posedness.  Specifically, the norm of the solution is not
bounded by the norm of the initial data irrespective of the data. A
long-running numerical experiment is performed as well, showing that
the type of ill-posedness observed may not be serious in specific
practical applications, as is known from many numerical simulations.

\end{abstract}

%\tableofcontents

%---------------------------------------------------------------------
\section{Introduction} \label{sec:1}

%---------------------------------------------------------------------

Taken at face value, the Einstein equations in their original form
are second-order equations for the metric components.  However,
rarely are they used in their original form for numerical
applications beyond the harmonic gauge. It is much more common to use
the 3+1 splitting, introducing the intrinsic and extrinsic curvatures
of a spacelike slice as the fundamental variables, in terms of which
the Einstein equations become a system of equations that are
first-order in time and second-order in space~\cite{york79}.  Some
variations are taken, such as decomposing the variables into
trace-free densities and the
traces~\cite{kurki-laguna-matzner93,shibata}, and, additionally,
partially converting the second-spatial derivatives to derivatives
of first-order variables~\cite{BS99}.  

Alhtough much has been learned in recent years about turning the 3+1
forms further down into a completely first-order
system~\cite{masso92,newtonian,masso95,york,moletter,helmut96,bspre,fixing},
little if anything has been said about these mixed first-in-time,
second-in-space forms so widely used. 

We investigate here the question of stability of some of these forms
under small changes of the initial data, namely, the question of
well-posedness.  We take a particular approach to this question,
based on quite standard norms for periodic solutions. The sense in
which the problems that we touch upon are said to be well-posed or
stable is the following. A system of equations for $m$ 
variables $u(x^i,t)$, $i=1,2,3$, would be stable if the norm of the
solution is bounded by the norm of the initial data in terms of
constants independent of the initial data, namely: 
\begin{equation}\label{generic}
||u(\cdot,t)||\le a e^{bt}
||f(\cdot)||,
\end{equation}
where $a,b$ are the same constants for all initial data
$f(x^i)=u(x^i,0)$. In the case of linear periodic problems, for
instance, this means that $a,b$ must be independent of the spectral
frequency of the initial data.  This definition is quite
standard~\cite{kreissbook}. Well-posedness in this or equivalent
sense is normally expected of physical systems, as well as of
successful approximation schemes, including numerical simulations.  

A problem for which the estimate (\ref{generic}) does not hold is
referred to as ill posed. While showing that the estimate holds for
all the solutions of a well-posed problem can be accomplished via
standard algebraic criteria, showing that the estimate does not hold
for a certain ill-posed problem may require no more than finding a
counterexample.  A suitable counterexample would consist of a
particular sub-family of solutions for which the estimate does not
hold.   

By systematically finding ``counterexamples'', we are able to
observe in Section~\ref{sec:2} that the standard ADM
form~\cite{york79} in the two most widely used gauges (geodesic and
harmonic slicings) is ill posed in the sense defined above, and that
its most recent conformally-decomposed version~\cite{BS99} is ill
posed as well.  We may consider this as a contribution to the 
direct analytic investigation of the standard ADM form, which some
authors~\cite{helmut96} have termed an outstanding problem.

This observation might cause a sudden loss of charm to these forms of
the Einstein equations. In this respect, it is definitely not our
intention to question the relevance of ill-posed problems in physics.
We refer the reader to page 230 of~\cite{courant}, where the
increasing importance of such problems is rightfully appreciated. 
After all, the main drawback of having to deal with an ill-posed
system is surely the current unavailability of relevant results from
mathematical physics.  

Far from questioning the use of the ADM forms in numerical
simulations, in Section~\ref{sec:3} we include a numerical simulation
of the standard ADM equations with geodesic slicing and periodic
boundaries.  We show that in spite of the ill-posed character, the
simulation can be carried out to long times with no signs of
instability apparent as yet.  Furthermore, we refer the reader
to~\cite{BS99} where numerical simulations with another equally
ill-posed version of the equations is integrated for long times with
no apparent instabilities.   We suggest that the ADM forms may retain
their charm, originating probably from the relatively small number of
fundamental variables and apparent compactness of the equations,
compared to fully first-order forms.  Nevertheless, it is likely
that numerical simulations of these equations might be suject to
their inherent instabilites, and an appropriate discretization for
long-time evolution may not suggest itself in an obvious manner. 

We conclude in Section~\ref{sec:4} with some remarks about the reach
and relevance of the results.

%---------------------------------------------------------------------
\section{Ill-posed ADM forms}\label{sec:2}

%---------------------------------------------------------------------

Because the reductions of the Einstein equations that we are interested in are
rather large systems, we first illustrate the procedure that we intend to use
in the much simpler case of the 1+1 wave equation. Our procedure is,
essentially, a rather systematic way of finding counterexamples to
(\ref{generic}).  We take it after an example in page 229 of~\cite{courant},
credited to Hadamard~\cite{hadamard}.  (It is hardly contestable that any
counterexample will suffice, irrespective of how particularly wicked!)

%---------------------------------------------------------------------
\subsection{The ill-posed form of the wave equation
\label{subsec:wave}}
%---------------------------------------------------------------------
 
Let's consider the
case of the 1+1 wave equation:
\begin{equation}
\ddot{\psi} = \psi_{xx}
\end{equation}
where $\dot{\,}\equiv \partial/\partial t$ and the sublabel $x$ stands for
$\partial/\partial x$.  Let's reduce the wave equation to
first-order-in-time, second-order-in-space form by defining a second
variable $\eta\equiv \dot{\psi}$.  We obtain thus a system
\begin{mathletters} \label{illwave}
\begin{eqnarray}
	\dot{\eta}&=&\psi_{xx}\\
	\dot{\psi}&=& \eta
\end{eqnarray}
\end{mathletters}
for a variable $u=\{\eta,\psi\}$ with
initial
values $f=\{\eta(x^j,0), \psi(x^j,0)\}$.  Let's consider, for
simplicity, 
the case of solutions of periodicity $L$.  A solution is 
\begin{eqnarray}
	\psi&=& \cos(\omega t)\sin(\omega x)\\
 	\eta&=& -\omega\sin(\omega t)\sin(\omega x)
\end{eqnarray}
where $\omega=2\pi n/L$ and $n$ is an integer. 
Different initial data are here labeled by different values of
$\omega$,
and so are the solutions arising from them, so we can think 
that at any given time and position, the value of the solution is a
function of the initial data through $\omega$.  Let's calculate the
norms:
\[
||f(\cdot)|| =\frac{1}{L} \int_0^L \psi(x,0)^2 + \eta(x,0)^2 dx 
             =\frac{1}{L} \int_0^L \sin^2(\omega x) dx
	     = \frac12 \hspace{1.5cm}\mbox{for all $\omega$} 
\]
and 
\begin{eqnarray}
||u(\cdot,t)|| 
&=&\frac{1}{L} \int_0^L \psi(x,t)^2 + \eta(x,t)^2 dx   \nonumber\\
&=&\frac{1}{L} ( \cos^2(\omega t) + \omega^2 \sin^2(\omega t))
		\int_0^L \sin^2(\omega x) dx	 \nonumber\\
&=& \frac12 ( \cos^2(\omega t) + \omega^2 \sin^2(\omega t))
\end{eqnarray}
so we have
\[
||u(\cdot,t)||
=\left( \cos^2(\omega t) + \omega^2 \sin^2(\omega
t)\right)||f(\cdot)||.
\]
Because $\omega^2$ has no upper bound in the full spectrum of
$\omega$,
there are no constants $a$ and $b$ such that $\left(
\cos^2(\omega t) + \omega^2 \sin^2(\omega t)\right)\le a e^{b
t}$ for all initial data (all $\omega$).  So we have a subset of
initial data for which no estimate holds, which is sufficient to
assert
that there is no estimate good for the entire set of data. This form
of the
wave equation is
ill posed.

This is sometimes interpreted as implying that a finite difference
scheme
for these equations would not progress forward in time without
instabilities even at short times, on the basis that it is virtually
impossible to filter out the high frequencies
$\omega$~\cite{kreissbook}. 

In the following, we examine two of the most commonly used versions
of the 3+1 splitting of the Einstein equations, namely the standard
ADM form, and the conformally-decomposed version that appeared
in~\cite{BS99}.  Because the latter is more involved, we develop it
first in some detail.

%---------------------------------------------------------------------
\subsection{Ill-posedness of the conformally-decomposed version of 
the ADM form
\label{subsec:bs}}
%---------------------------------------------------------------------

The system of interest has appeared in~\cite{BS99}, and is based on
earlier work in~\cite{shibata}. It is a system of 15 equations for 15
variables $(\phi, K, \tilde{\gamma}_{ij}, \tilde{A}_{ij},
\tilde{\Gamma}^i)$, and is referred to as System II in~\cite{BS99},
to distinguish it from the standard 3+1 Einstein
equations~\cite{york79}, which we display in the next subsection,
Eqs.~(\ref{admeqs}).  These variables are related to the
intrinsic metric $\gamma_{ij}$ and extrinsic curvature $K_{ij}$ as
follows:
\begin{mathletters}
\begin{eqnarray}
	e^{4\phi} &=& \det(\gamma_{ij})^{1/3}\\
	\tilde{\gamma}_{ij} & = & e^{-4\phi}\gamma_{ij}\\
	K  &=& \gamma^{ij}K_{ij}\\
	\tilde{A}_{ij} &=& e^{-4\phi}\left( 
				K_{ij} 
		  - \frac13 \gamma_{ij}K\right)	\\
	\tilde{\Gamma}^i &=& -\tilde{\gamma}^{ij},_j
\end{eqnarray}
\end{mathletters}
where $\tilde{\gamma}^{ij}$ is the inverse of $\tilde{\gamma}_{ij}$.
The Einstein evolution equations (\ref{admeqs}) in terms of these
variables are equivalent~\cite{BS99} to the following:
\begin{mathletters}\label{bs}
\begin{eqnarray}
  \frac{d}{dt}\phi &=& -\frac16 \alpha K\\
  \frac{d}{dt}\tilde{\gamma}_{ij}& = & -2\alpha\tilde{A}_{ij}\\
  \frac{d}{dt}K  				&=& 
	-\gamma^{ij}D_iD_j\alpha 
	+\alpha\left(\tilde{A}_{ij}\tilde{A}^{ij}
		     +\frac13 K^2
		\right)
	+\frac12\alpha(\rho+S)\\
  \frac{d}{dt}\tilde{A}_{ij} 			&=& 
	e^{-4\phi}\left( -(D_iD_j\alpha)^{TF}
			+\alpha(R_{ij}^{TF} - S_{ij}^{TF})
		  \right)
	+\alpha(K\tilde{A}_{ij}-2\tilde{A}_{il}\tilde{A}^l{}_j) \\
  \frac{\partial}{\partial t}\tilde{\Gamma}^i 	&=& 
	-2\tilde{A}^{ij}\alpha,_j
	+2\alpha\left(\tilde{\Gamma}^i_{jk}\tilde{A}^{kj}
		      -\frac23\tilde{\gamma}^{ij}K,_j
		      -\tilde{\gamma}^{ij}S_j
		      +6\tilde{A}^{ij}\phi,_j
		\right)					\nonumber\\
&&	-\frac{\partial}{\partial x^j}
	 \left( \beta^l\tilde{\gamma}^{ij},_l
		-2\tilde{\gamma}^{m(j}\beta^{i)},_m
		+\frac23\tilde{\gamma}^{ij}\beta^l,_l
	 \right).
\end{eqnarray}
\end{mathletters}
Here $\alpha$ is the lapse function, $\beta^i$ is the shift vector,
and $\tilde{\Gamma}^i_{jk}$ are the connection coefficients of
$\tilde{\gamma}_{ij}$. The superscript $^{TF}$ denotes trace-free
part, e.g. $R_{ij}^{TF} = R_{ij} - \gamma_{ij}R/3$. Indices are
raised and lowered with $\tilde{\gamma}^{ij}$ and its inverse. We use
the shorthand notation
\begin{equation}
\frac{d}{dt} \equiv \frac{\partial}{\partial t} - \pounds_\beta
\end{equation}
where $\pounds_\beta$ is the Lie derivative along $\beta^i$. Although
for the solution to (\ref{bs}) to be a solution to the full set of
Einstein equations it is necessary that the initial data satisfy the
set of four additional constraints, in the following, we do not need
to consider the four constraint equations explicitly, since they do
not affect the well-posedness of the evolution system.  We
assume that the constraints are imposed on the initial data, thus
selecting, from the set of solutions to (\ref{bs}), the subset that
satisfies the ten Einstein equations.

The system (\ref{bs}) is first-order in time and second-order in space,
being generically represented in the form 
\begin{equation} \label{gensys}
\dot{u} = 
	  A^{ij}(x^k,t,u)u,_{ij} 
	+ B^i(x^k,t,u,u_{,k}) u,_i 
	+ C(x^k,t,u)
	\equiv P(x^k,t,u,u_{,k},\partial/\partial x^j)u. 
\end{equation}
Any time that an evolution system can be represented in
this form, we interpret the right-hand side as an evolution
operator acting on a solution. In this case, the evolution operator
is $P$, and contains all the terms in the right-hand side of
(\ref{bs}).  It is essential to point out that the properties of
stability of a system like (\ref{gensys}) are encoded in the
principal terms of the operator $P$, namely, in
$A^{ij}(x^i,t,u)$. This means that we can restrict our attention
to  a system of the form
\begin{equation}\label{genprincipal}
\dot{u} = 
	  A^{ij}(x^i,t,u)u,_{ij}. 
\end{equation}
The lower-order terms that differentiate $A^{ij}(x^i,t,u)$ from
$P$ do not affect the existence of an estimate of the form
(\ref{generic}) (see, for instance, p. 139 of~\cite{kreissbook}).  If
there is such an estimate for (\ref{genprincipal}), then there is one
for (\ref{gensys}) as well, and for any other system with a
right-hand-side that differs from $P$ only in first and zeroth order
terms.  If there is no estimate for (\ref{genprincipal}), then there
is no estimate for (\ref{gensys}) either, nor for any system that
differs from (\ref{bs}) in first-derivatives or undifferentiated
terms. 

In the following, we focus on the principal terms of the evolution
operator of (\ref{bs}), namely, the exact terms that contain the
highest derivatives of the dynamical fields in the right-hand-side.
In this case, the principal terms are
\begin{mathletters}\label{principal}
\begin{eqnarray}
  \dot{\phi} &=& 0,\\
  \dot{\tilde{\gamma}_{ij}}& = & 0,\\
  \dot{K} 		&=& 0,\\
  \dot{\tilde{A}_{ij}} 			&=& 
	e^{-4\phi}\alpha\left(
	-\frac12 \tilde{\gamma}^{lm}\tilde{\gamma}_{ij,lm}
	-2\left(\phi,_{ij}
	        -\frac13\tilde{\gamma}_{ij}
			\tilde{\gamma}^{lm}\phi,_{lm}
	   \right)	  
			\right),\label{adot}\\
  \dot{\tilde{\Gamma}^i} &=&
  \beta^l\tilde{\gamma}^{ij},_{jl},		\label{bsGamma}
\end{eqnarray}
\end{mathletters}
where $\dot{\;} \equiv \partial/\partial t$.  Second-derivatives of
the lapse and shift do not contribute to the principal part of the
evolution operator of (\ref{bs}) if they are assumed to be
arbitrarily given as source functions.  If they were given
dynamically, in terms of the metric or extrinsic curvature, then
their second-derivatives would contribute to the principal part of the
evolution operator and should be included (as an example, see the
case of the harmonic slicing below).  

For our purposes, it is much simpler to work with system
(\ref{principal}) than with (\ref{bs}) without affecting our
conclusions. The system (\ref{principal}) differs from (\ref{bs}), but
only in terms that are of first and zeroth order. Clearly the solutions
will be different, but not their stability properties.  It may not be
obvious that the terms that are essential to make (\ref{adot})
trace-free are of first-order (not second), and therefore do not need to
be included in our discussion.  The reader may as well consider them as
included, since their inclusion does not affect our argument in any
way. 

We can show that (\ref{principal}) is ill posed by finding one family of
solutions for which no estimate of the form (\ref{generic}) holds. 
Essentially the same procedure that gave us the result for the wave
equation gives the analogous result for (\ref{principal}), and
consequently, for (\ref{bs}).  However, because the procedure consists
of finding explicit sloutions, the results are necessarily restricted by
the gauge.  We will determine ill-posedness in two cases: geodesic and
harmonic slicing.  For geodesic slicing, assume $\alpha = 1$ and
$\beta^i=0$.  Consider the periodic solution
\begin{mathletters} 
\begin{eqnarray}
\phi&=&\frac14\log\left(1+\frac12\cos(k\!\cdot\! x)\right),\\
\tilde{\gamma}_{ij}&=&\delta_{ij},\\
	K&=&0,\\
\tilde{A}_{ij} &=& \frac{\left(\frac12 + \cos(k\!\cdot\! x)\right)}
			{4\left(1+\frac12 \cos(k\!\cdot\! x)\right)^3}
		\left(k_ik_j
		      -\frac13 k\!\cdot\! k\, \delta_{ij}\right) t,\\
\tilde{\Gamma}^i &=&0,
\end{eqnarray}
\end{mathletters}
where $k\!\cdot\! x \equiv k^ix^j\delta_{ij}$, and $k^i = 2\pi n^i/L$
with integer $n^i$.  The initial data for this solution are
\begin{mathletters} 
\begin{eqnarray}
\phi(x,0)&=&\frac14\log\left(1+\frac12\cos(k\!\cdot\! x)\right),\\
\tilde{\gamma}_{ij}(x,0)&=&\delta_{ij},\\
	K(x,0)&=&0,\\
\tilde{A}_{ij}(x,0) &=& 0,\\
\tilde{\Gamma}^i(x,0)&=&0. 
\end{eqnarray}
\end{mathletters}
The norm is defined as
\begin{equation}\label{L2norm}
||u(\cdot,t)|| 
   \equiv 
	\frac{1}{L^3}\int_{cube}\!\!d^3x\;
	\Big(
	 \phi^2
	+\sum_{ij}(\tilde{\gamma}_{ij})^2
	+\sum_i(\tilde{\Gamma}^i)^2
	+K^2
	+\sum_{ij}(\tilde{A}_{ij})^2
	\Big).
\end{equation} 
Let's calculate the norms at the initial time and at time $t$. We have
\begin{equation}
||u(\cdot,0)||\equiv ||f(\cdot)||
   =
   3 + \frac{1}{L^3}\int_{cube}
	\left( \frac14\log\left(1+\frac12\cos(k\!\cdot\! x)\right)
	\right)^2 d^3x,
\end{equation}
and 
\begin{eqnarray}
||u(\cdot, t)||
   &=&
   3 + \frac{1}{L^3}\int_{cube}
	\left( \frac14\log\left(1+\frac12\cos(k\!\cdot\! x)\right)
	\right)^2 d^3x			\nonumber\\
  && + \frac{1}{L^3}\int_{cube}
	\frac23 (k\!\cdot\! k)^2 t^2
	\left( \frac{\cos(k\!\cdot\! x)+\frac12}
		    {4(1+\frac12 \cos(k\!\cdot\! x))^3}
	\right)^2 d^3x.
\end{eqnarray}
It is our purpose to demonstrate that the norm $||u(\cdot, t)||$ can not
be bounded independently of $\omega\equiv\sqrt{k\!\cdot\! k}$.  For this
purpose, we start by factoring out $||f(\cdot)||$, namely:
\begin{equation}  \label{step1}
||u(\cdot, t)||
   = 
	||f(\cdot)||
 	\left(
	1
   +    \frac23 \frac{(k\!\cdot\! k)^2 t^2}{16 ||f(\cdot)||}
	\frac{1}{L^3}\int_{cube}
	\left( \frac{\cos(k\!\cdot\! x)+\frac12}
		    {(1+\frac12 \cos(k\!\cdot\! x))^3}
	\right)^2 d^3x
	\right).
\end{equation}
Because $1+\frac12 \cos(k\!\cdot\! x) \le 3/2$, we have that
\begin{equation}
\left( \frac{\cos(k\!\cdot\! x)+\frac12}
	    {(1+\frac12 \cos(k\!\cdot\! x))^3}
\right)^2
	\ge
\left( \left(\frac23\right)^3 \left(\cos(k\!\cdot\! x)+\frac12\right)
\right)^2,
\end{equation}
and because $\int_{cube}(\cos(k\!\cdot\! x)+\frac12)^2 d^3x = 3L^3/4$
then
\begin{equation}
\frac{1}{L^3}\int_{cube} 
\left( \frac{\cos(k\!\cdot\! x)+\frac12}
	    {1+\frac12 \cos(k\!\cdot\! x)}
\right)^2 d^3x
	\ge
\frac34 \left(\frac23\right)^6.
\end{equation}
Using this inequality into (\ref{step1}) we obtain
\begin{equation}  \label{step2}
||u(\cdot, t)||
   \ge 
	||f(\cdot)||
 	\left(
	1
   +    \frac23 \frac{(k\!\cdot\! k)^2 t^2}{16 ||f(\cdot)||}
	\frac34\left(\frac23\right)^6
	\right).
\end{equation}
Also because $1+ \frac12 \cos(k\!\cdot\! x)\le 3/2$, we have that
\begin{equation}
 \frac{1}{||f(\cdot)||} 
   \ge 
	\frac{1}{3+\left({\displaystyle \frac{\log(3/2)}{4}}\right)^2}
\end{equation}
which, if plugged into (\ref{step2}), yields
\begin{equation}
||u(\cdot, t)||
   \ge 
	||f(\cdot)||
 	\left(
	1
   +    \frac23 \frac{(k\!\cdot\! k)^2 t^2}{16}
		\frac{1}{3+\left({\displaystyle \frac{\log(3/2)}{4}}
			   \right)^2}
	\frac34\left(\frac23\right)^6
	\right).
\end{equation}
Because $(k\!\cdot\! k)^2$ is unbounded, then $||u(\cdot, t)||$
increases out of bound at large frequencies $\omega$. This shows that an
estimate of the type (\ref{generic}) does not exist.  Therefore the
system (\ref{principal}) is ill posed.  The addition of
first-derivatives or of undifferentiated terms will not turn
(\ref{principal}) into a well-posed system, from which it follows that
(\ref{bs}) is ill posed as well. The argument may be impossible to
generalize to the case of arbitrary lapse and shift, but presently it
suffices to make our point. 

Consider now the harmonic slicing $\alpha = \sqrt{\det\gamma_{ij}}$ with
$\beta^i=0$.  In this case, the principal terms of the evolution
operator of the system (\ref{bs}) are not (\ref{principal}), because the
second-order derivatives of the lapse are now dynamical and must be
considered. We actually have 
$\alpha = e^{6\phi}$ and $
	\alpha,_{ij} = 6 e^{6\phi} (\phi,_{ij} + 6\phi,_i\phi,_j)$.
Thus the principal terms of the evolution of (\ref{bs}) in the harmonic
gauge are
\begin{mathletters}\label{principalharm}
\begin{eqnarray}
  \dot{\phi} &=& 0,\\
  \dot{\tilde{\gamma}_{ij}}& = & 0,	\\
  \dot{K} 		&=&  
			   -6 e^{2\phi}	\tilde{\gamma}^{kl}\phi,_{kl},\\
  \dot{\tilde{A}_{ij}} 			&=& 
  	e^{2\phi}\left(
	-\frac12 \tilde{\gamma}^{lm}\tilde{\gamma}_{ij,lm}
	-8\left(\phi,_{ij}
	-\frac13\tilde{\gamma}_{ij}\tilde{\gamma}^{lm}\phi,_{lm}
	  \right)
			  \right),				\\
  \dot{\tilde{\Gamma}^i} &=&
  \beta^l\tilde{\gamma}^{ij},_{jl},
\end{eqnarray}
\end{mathletters}
where $\dot{\;} \equiv \partial/\partial t$. A periodic solution is
\begin{mathletters}
\begin{eqnarray}
	\phi&=&\frac12\log\left(1+\frac12\cos(k\!\cdot\! x)\right), \\
	 \tilde{\gamma}_{ij} 
	     &=& \delta_{ij}					,\\
	K   &=&6 t  
		   \frac{\left(\frac12 + \cos(k\!\cdot\! x)\right)}
			{\left(1+\frac12 \cos(k\!\cdot\! x)\right)}
			k\!\cdot\! k 				,\\
	\tilde{A}_{ij} &=&8 t 
			\frac{\left(\frac12 + \cos(k\!\cdot\! x)\right)}
			{\left(1+\frac12 \cos(k\!\cdot\! x)\right)}
		\left(k_ik_j-\frac13 k\!\cdot\! k\, \delta_{ij}\right),\\  			
	\tilde{\Gamma}^i&=&0.
\end{eqnarray}
\end{mathletters}
With the norm (\ref{L2norm}) and by following very similar calculations
as in the case of geodesic slicing, one can readily see that the
following inequality holds:
\begin{equation}
||u(\cdot, t)||
   \ge 
	||f(\cdot)||
 	\left(
	1
   +   (k\!\cdot\! k)^2 t^2
	\left(\frac23\right)^2
	\frac{59}{3+\left({\displaystyle \frac{\log(3/2)}{2}}\right)^2}
	\right).
\end{equation}
As in the previous cases, we can see that the norm is not bounded
because of the presence of $\omega^4$.   This shows that the harmonic
slicing does not ``turn'' system (\ref{bs}) into a well-posed form.  This
appears to contradict standard theorems on the
well-posedness of the Einstein equations in the harmonic gauge. 
However, it does not. The difference between the use of the harmonic
gauge in standard proofs of well-posedness and the treatment in the
present paper is that the standard proofs are carried out with the
Einstein equations either in full first-order form or in the original
second-order form, whereas here we are considering the mixed case of
first-order in time and second-order in space.

%---------------------------------------------------------------------
\subsection{Ill-posedness of the standard ADM form
\label{subsec:adm}}
%---------------------------------------------------------------------

We now consider the standard ADM equations in the form
proposed in~\cite{york79}, which are of wide use in numerical
applications. The equations are
\begin{mathletters}
\label{admeqs}
\begin{eqnarray}
	\dot{\gamma}_{ij}
   &=&
	-2\alpha K_{ij} + 2D_{(i}\beta_{j)}		\\
	\dot{K}_{ij},				\label{admgammadot}
   &=&
\label{admkdot}
	\alpha R_{ij}
	-2\alpha K_{ik}K_j^k
	+\alpha K K_{ij}
	-D_iD_j \alpha
	+\beta^kD_kK_{ij}
	-2K_{k(i}D^k\beta_{j)},
\end{eqnarray}
\end{mathletters}
where $\gamma_{ij}$ is the intrinsic metric of the slice at constant
$t$, $K_{ij}$ is the extrinsic curvature of the slice, defined by
(\ref{admgammadot}), $\alpha$ is the lapse function and $\beta^i$ is
the shift vector.  Here we also benefit from restricting attention to
the principal terms in the right-hand-side, namely the exact
second-derivative terms. 
Assuming non-dynamical choices of lapse and shift (ruling out the
harmonic slicing, in particular), the principal terms of the evolution
operator of the system are
\begin{mathletters}\label{admprincipal}
\begin{eqnarray}
	\dot{\gamma}_{ij}
   &=&
	0,	\\
	\dot{K}_{ij}
   &=&
	\frac{\alpha}{2} 
	\left(
		2\gamma^{kl}\gamma_{l(i,j)k}
		-\gamma^{kl}\gamma_{ij,kl}
		-\gamma^{kl}\gamma_{kl,ij}
	\right)	.
        \label{kdot}
\end{eqnarray}
\end{mathletters}
Consider the following periodic solution to (\ref{admprincipal}) with
$\alpha=1$ (geodesic slicing):
\begin{mathletters}
\begin{eqnarray}
\gamma_{ij} &=& \left(1+\frac12 \cos(k\!\cdot\! x)
		\right) \delta_{ij},	\\
K_{ij}      &=& t\frac{ \cos(k\!\cdot\! x)}
		      {\left(1+\frac12 \cos(k\!\cdot\! x)
		       \right)}
		\left(k_ik_j+\delta_{ij}k\!\cdot\! k \right).
\end{eqnarray}
\end{mathletters}
With the norm
\begin{equation}\label{admL2norm}
||u(\cdot,t)|| 
   \equiv 
	\frac{1}{L^3}\int_{cube}\!\!d^3x\;
	\Big(
	\sum_{ij}(\gamma_{ij})^2
	+\sum_{ij}(K_{ij})^2
	\Big),
\end{equation}
we have
\begin{equation}
 ||u(\cdot,t)||
   =
	||f(\cdot)|| 
	+ 6 (k\!\cdot\! k)^2 t^2 
	\frac{1}{L^3}\int_{cube}\!\!d^3x\;
		\frac{\cos^2(k\!\cdot\! x)}
		     {(1+\frac12 \cos(k\!\cdot\! x))^2},
\end{equation}
and
\begin{equation}
 ||f(\cdot)||
   = 
	\left(\frac32 \right)^3.
\end{equation}
It is very simple to see that the following inequality holds:
\begin{equation}
||u(\cdot, t)||
   \ge 
	||f(\cdot)||
 	\left(
	1
   +    \frac{2^5}{3^4}
	(k\!\cdot\! k)^2 t^2
	\right)
\end{equation}
Because of the presence of $\omega^4$, we conclude that the norm is
not bounded in terms of the initial data and the standard ADM
equations are ill-posed for $\alpha=1$, in the same sense as system
(\ref{bs}) is.

For completeness, we end this section by showing that even the
harmonic gauge suffers from the same type of ill-posedness.  For the
harmonic gauge we have $\alpha=\sqrt{\det\gamma_{ij}}$, thus
$D_iD_j\alpha=(\alpha/2)\gamma^{kl}\gamma_{kl,ij}+\cdots$, so that the
principal terms of (\ref{admeqs}) are
\begin{mathletters}\label{admharmprincipal}
\begin{eqnarray}
	\dot{\gamma}_{ij}
   &=&
	0	,\\
	\dot{K}_{ij}
   &=&
	\frac{\alpha}{2} 
	\left(
		2\gamma^{kl}\gamma_{l(i,j)k}
		-\gamma^{kl}\gamma_{ij,kl}
		-2\gamma^{kl}\gamma_{kl,ij}
	\right).	
\end{eqnarray}
\end{mathletters}
Consider the following periodic solution to (\ref{admharmprincipal}):
\begin{mathletters}
\begin{eqnarray}
\gamma_{ij} &=& \left(1+\frac12 \cos(k\!\cdot\! x)
		\right) \delta_{ij}	,\\
K_{ij}      &=& \frac{t}{4}
	\left(1+\frac12 \cos(k\!\cdot\! x)\right)^{\frac12}
		 \cos(k\!\cdot\! x)
		\left(4k_ik_j+\delta_{ij}k\!\cdot\! k \right).
\end{eqnarray}
\end{mathletters}
With the norm (\ref{admL2norm}) it is straightforward to see that the
following inequality holds:
\begin{equation}
||u(\cdot, t)||
   =
	||f(\cdot)||
 	\left(
	1
   +    \frac{5}{18}
	(k\!\cdot\! k)^2 t^2
	\right).
\end{equation}
It is impossible to bound the terms in parenthesis in the
right-hand-side by a factor of the form $ae^{bt}$ with $a$ and $b$
independent of $\omega=\sqrt{k\!\cdot\! k}$. Thus an estimate of the
form (\ref{generic}) does not hold, and the standard ADM equations in
the harmonic gauge are ill posed.

%---------------------------------------------------------------------
\section{Numerical experiments }
\label{sec:3}

%---------------------------------------------------------------------

We have seen in previous sections that the standard ADM form of the
Einstein equations is ill-posed in a certain definite sense, and that
so are the conformally-decomposed version (\ref{bs}) and even the 1+1
wave equation in flat space, against maybe widespread intuition.
Although this may create a difficulty in the stability analysis of
these systems, we do not think that this necessarily implies an
obstruction to numerical integration of any of them, in general.  For
particular applications, it is possible that the bad behavior at high
frequencies may be disregarded. On the other hand, there may be
discretizations suitable for numerical evolution for long enough
times, not necessarily arbitrarily long.  

In this section we present two numerical experiments with these
ill-posed systems.  In the first place, we show that in the case of
the ill-posed form of the 1+1 wave equation there exists a stable
discretization.  This in principle implies that the numerical
integration is not hampered in any way. The fact that a stable
discretization exists in this case in spite of the ill-posedness is
very unusual;  it is possible that the 1+1 wave equation be an
exception to the rule. 

Secondly, we present a discretization of the standard ADM equations
in the geodesic slicing which runs for impressively long times
without any signs of instabilities.

%---------------------------------------------------------------------
\subsection{Stable discretization for the ill-posed wave equation
\label{subsec:wavenum}}

%---------------------------------------------------------------------

In this subsection we discretize the ill-posed form of the wave
equation, namely (\ref{illwave}).  This is a standard exercise, but a
very illuminating one. Our intention is to emphasize that the
discretization is stable, which means that the discretized equations
have no modes that are amplified by the evolution. This is in spite
of the ill-posed character of the continuous equations that the
discretization is modeling.

We use the {\it leapfrog\/} discretization, which is explicitly given by
\begin{mathletters}
\begin{eqnarray}\label{discrete}
	\eta^{n+1}_j  &=& \frac{2\,\Delta t}{h^2}
			\left(  \psi^n_{j+1} 
			      -2\psi^n_j
			      + \psi^n_{j-1}
			\right)
			+ \eta^{n-1}_j\\
	\psi^{n+1}_j &=& 2\,\Delta t \eta^n_j 
			+ \psi^{n-1}_j
\end{eqnarray}
\end{mathletters}   
where $\Delta t$ represents the time step, and $h$ represents the spacing
between the grid points, namely $h=\Delta x$. The {\it grid} points
are equally spaced in the interval $-1\le x \le 1$, $x_i=-1+(i-1)h$
for $i=1\ldots N$ and $h=2/(N-1)$.

In order to carry out a stability analysis~\cite{recipes}, we use 
the ansatz 
\begin{eqnarray}
	\eta^n_j = \xi^n e^{ijkh} \eta_0 \\
	\psi^n_j = \xi^n e^{ijkh} \psi_0
\end{eqnarray}
where $i\equiv\sqrt{-1}$.  Substituting this ansatz into
(\ref{discrete}), we obtain 
\begin{equation}
\left(
\begin{array}{cc}
	(\xi^2-1) & -4\frac{\Delta t}{h^2} (\cos(kh)-1)\xi	\\
	-2\,\Delta t \xi & (\xi^2-1)
\end{array}
\right)
\left(
\begin{array}{c}
\eta_0	\\
\psi_0
\end{array}
\right)
  =
     0
\end{equation}
For a non-trivial solution we need the determinant of the system to be
zero, namely
\begin{equation}
	(\xi^2-1)^2 + 2 \alpha^2 (1-\cos(kh))\xi^2 = 0
\end{equation}
where $\alpha \equiv \frac{2\,\Delta t}{h}$. The solution is 
\begin{equation}
	\xi^2 = -(c-1) \pm \sqrt{(c-1)^2 - 1}
\end{equation}
with $c\equiv \alpha^2(1-\cos(kh))$.  The discretization is stable if
$|\xi| \le 1$.  A sufficient condition for this to happen in our case is
that $c \le 2$, where we have $\xi\bar{\xi} = 1$, i.e. the discretization
is not only stable but unimodular. The requirement for $c\le 2$ is that
$\alpha \le 1$.  This means that for our discretization to be stable and
unimodular we only need to take the time step smaller than one-half the
grid size $h$.

\begin{figure}
\centerline{\epsfysize=4in\epsfbox{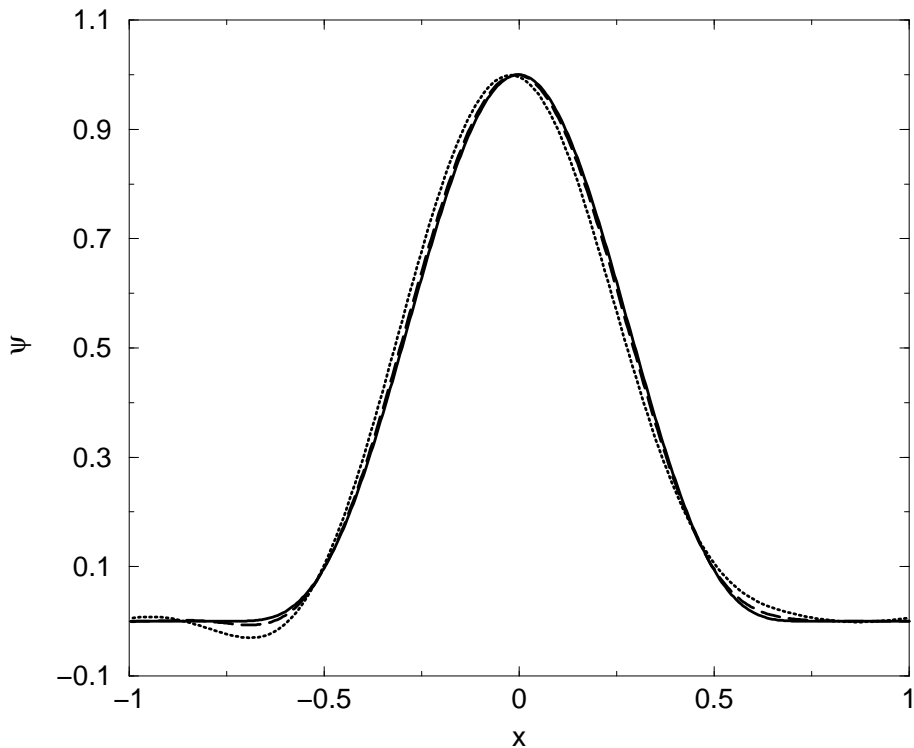}}
\caption{Evolution of a pulse traveling to the right at speed
$v=1$. Shown in this figure are the initial $(t=0.0)$ and the final pulse
$(t=200.0)$. The later time is seen at the two grid resolutions, $N=200$
(dotted line) and $N=400$ (dashed line).}
\label{wavefig}
\end{figure}

We implemented a straightforward Fortran code based upon the discretization
(\ref{discrete}), with periodic boundary conditions in the domain $-1\le
x\le 1$. As expected, the code reproduces very well the evolution
of a pulse of compact support of the form 
\begin{equation} \psi = \left\{
\begin{array}{cc}
\displaystyle{\frac{A}{w^8}}((t - v x)^2 - w^2) &
  {\rm for} \quad |t - v x|\le w , \\
0 & {\rm otherwise} ,
\end{array}
\right.
\end{equation}
where the evaluation of the condition is to be understood modulo 2 (the 
periodicity of the grid). The pulse shown in Fig.~\ref{wavefig} has amplitude
$A=1$ and width $w=1$, and it is propagating to the right with speed
$v=1$.  We can see no sign of damping or amplitude growth, and very
little distortion, even after 100 crossing times.  Shown in the figure
are the initial time, $t=0.0$ (solid line) and the final pulse,
$t=200.0$, at a resolution of $N=200$ points (dotted line) and $N=400$
points (dashed line) respectively. Note how the distortion of the pulse
decreases with increasing resolution as expected from the second-order
discretization (\ref{discrete}).

%---------------------------------------------------------------------
\subsection{Numerical stability of the ADM
equations\label{subsec:admnum}}
%---------------------------------------------------------------------

The ADM equations in the form introduced by York~\cite{york79}, and variations
of the same, have been used extensively in numerical
work~\cite{kurki-laguna-matzner93,shibata,BS99,admlong1,admlong2,admlong3,admlong4},
to cite a few. For completeness, however, we present here an alternate
straightforward numerical implementation of the ADM equations, which
illustrates that one can indeed integrate the nonlinear system (not just its
linearized counterpart, and not just its principal part). Starting from
(\ref{admeqs}), we restrict our attention to the case of geodesic slicing,
where $\alpha=1$, $\beta^i=0$. The calculation of the Ricci tensor in
(\ref{admkdot}) is the most involved and error-prone task, but from the
standard definition~\cite{wald84}
\begin{mathletters}
\begin{eqnarray}
  R_{ij} &=& - \Gamma^{k}_{ij,k} + \Gamma^{k}_{ik,j} 
  - \Gamma^{n}_{im} \Gamma^{m}_{jn} + \Gamma^{n}_{ni} \Gamma^{m}_{mj},
\label{ricci} \\
  \Gamma^{k}_{ij} &=& \frac{1}{2} \gamma^{kl} 
  \left( - \gamma_{ij,l} + \gamma_{jl,i} + \gamma_{li,j} \right),
\label{chris2}
\end{eqnarray}
\end{mathletters}
it follows that its computation can be organized by collecting terms
which contain second derivatives, terms with first derivatives of
the metric and its inverse, and terms with only connection
components,
\begin{eqnarray}
   R_{ij} &=& \frac{1}{2} \gamma^{kl}
       \left( 2\gamma_{l(i,j)k} -\gamma_{ij,kl} -\gamma_{kl,ij} \right)
 \nonumber \\
          && +\frac{1}{2} \gamma^{kl}{}_{,k} 
   \left(- \gamma_{ij,l} + \gamma_{jl,i} + \gamma_{li,j} \right)
 - \frac{1}{2} \gamma^{kl}{}_{,i} 
   \left(- \gamma_{kj,l} + \gamma_{jl,k} + \gamma_{lk,j} \right)
 \nonumber \\
               &&
  - \Gamma^{n}_{im} \Gamma^{m}_{jn} + \Gamma^{n}_{ni} \Gamma^{m}_{mj}.
\label{dricci}
\end{eqnarray}
We first compute the inverse metric and the second spatial
derivatives of the metric, from which we obtain the first term, then
calculate the first derivatives of the metric (and of the inverse
metric), which yields the second and third term. The final step is to calculate
the connection coefficients~(\ref{chris2}) and the last two terms in
(\ref{dricci}). The terms involving $K_{ij}$ in the right hand side
of~(\ref{admeqs}) are expressed in terms of the (downstairs)
extrinsic curvature and the inverse metric,
\begin{equation}
   K K_{ij} -2 K_{ik}K_j^k = \gamma^{lm} 
   \left(K_{lm} K_{ij} - 2 K_{li} \, K_{mj} \right).
\end{equation}

We discretize (\ref{admeqs}) using (\ref{chris2})-(\ref{dricci}),
taking the variables $\gamma_{ij}$ and $K_{ij}$ as given on a
rectangular, equally spaced grid of size $N^3$, with resolution
$h=2\pi/N$ and covering the domain $0\le x\le2\,\pi$, $0\le
y\le2\,\pi$ and $0\le z\le2\,\pi$. We label the grid points by
$x_k=k\,h$, $y_l=l\,h$, $z_m=m\,h$, with $k,l,m=1\ldots N$, and the time
levels as $t=t_0+n\Delta t$, for $n=1\ldots N_t$. To distinguish tensor
indices from grid indices, we use the notation
$\gamma_{ij}[{}^{n}_{k,l,m}]=\gamma_{ij}(x_k,y_l,z_m,t^n)$. We use
symmetry where appropriate, storing only the relevant components of
$\gamma_{ij}$, $K_{ij}$, $\gamma^{ij}$, $\gamma_{ij,k}$,
$\gamma^{ij}{}_{,k}$, $\gamma_{ij,kl}$ and $R_{ij}$. For instance, in the
calculation of the second and third terms of (\ref{dricci}), the
$\gamma^{ij}{}_{,k}$ are stored on the space allocated later to the
$\Gamma_{ij}^{k}$.

We compute first spatial derivatives $\gamma_{ij,k}$ and
$\gamma^{ij}{}_{,k}$, and second spatial derivatives $\gamma_{ij,kl}$
with centered, second-order accurate finite differences on grid
points, e.g.
\begin{mathletters}
\begin{eqnarray}
   \psi_{,x}[{}^{n}_{i,j,k}] &=& \frac{1}{2h} 
      \left( \psi[{}^{n}_{i+1,j,k}] - \psi[{}^{n}_{i-1,j,k}] \right) 
\\
   \psi_{,xx}[{}^{n}_{i,j,k}] &=& \frac{1}{h^2} 
      \left( \psi[{}^{n}_{i+1,j,k}] - 2\, \psi[{}^{n}_{i,j,k}] 
           + \psi[{}^{n}_{i-1,j,k}] \right) 
\\
   \psi_{,xy}[{}^{n}_{i,j,k}] &=& \frac{1}{4h^2}
      \left( \psi[{}^{n}_{i+1,j+1,k}] - \psi[{}^{n}_{i-1,j+1,k}] 
           - \psi[{}^{n}_{i+1,j-1,k}] + \psi[{}^{n}_{i-1,j-1,k}] 
      \right) 
\end{eqnarray}
\end{mathletters}
We perform the grid indices operations identifying the points $x_0
\equiv x_N$ and $x_{N+1} \equiv x_1$, which enforces periodic boundary
conditions in all coordinates.

The time integration scheme we use is the so-called iterative
Crank-Nicholson (ICN) method~\cite{admlong1,admlong2,BS99}, with three iterations
(i.e. one predictor step ``forward in time'', plus two correction
steps). Considering~(\ref{admeqs}) as equations of the form $ \dot{u}
= P u$, where $P$ is an operator acting on $u=(\gamma_{ij},K_{ij})$,
the time integration algorithm is given by
\begin{mathletters}
\begin{eqnarray} 
  u_{1}^{n+1} &=& u^{n} + \Delta t \, P u^{n}  
   \\
  \nonumber \\
  u_{2}^{n+\frac{1}{2}} &=& 
\frac{1}{2} \left( u_{1}^{n+1} + u^{n} \right)
   \\
  u_{2}^{n+1} &=& u^{n} + \Delta t \, P u_{2}^{n+\frac{1}{2}}
   \\
  \nonumber \\
  u_{3}^{n+\frac{1}{2}} &=& \frac{1}{2} \left( u_{2}^{n+1} + u^{n} \right)
   \\
  u^{n+1} &=& u^{n} + \Delta t \, P u_{3}^{n+\frac{1}{2}}
\end{eqnarray}
\end{mathletters} 
where quantities with sub-indices are intermediate values which do
not require additional storage.

\begin{figure}
\centerline{\epsfysize=4in\epsfbox{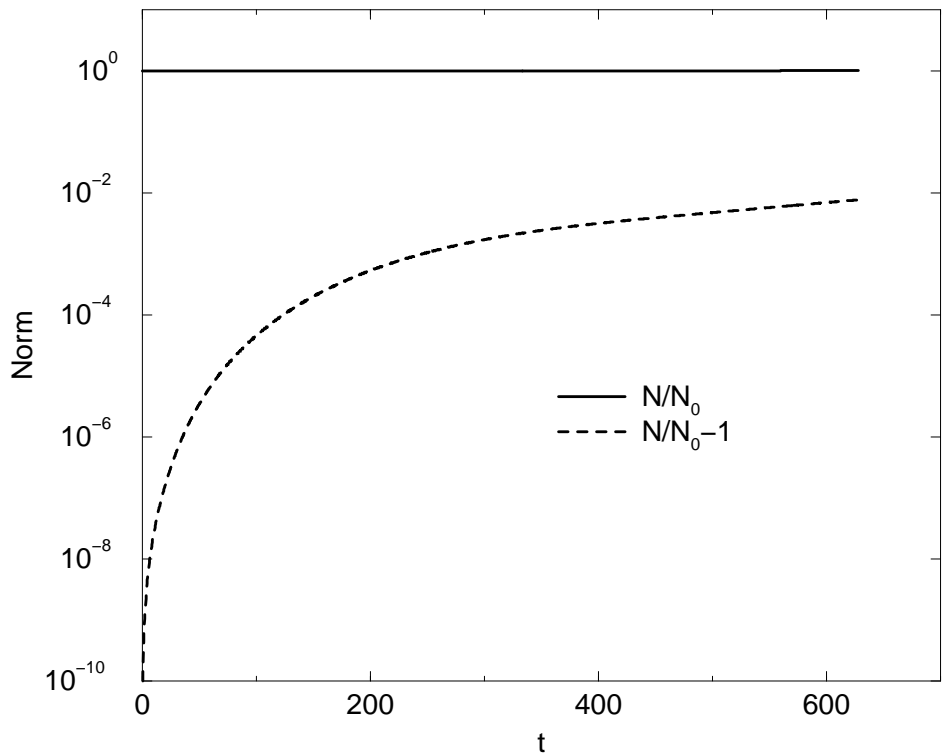} }
\caption{Evolution of a linearized, periodic ADM pulse traveling in
the direction $k=(1,2,1)$. Shown in this figure are the scaled norm
$N/N_0\equiv ||u(\cdot,t)||/||f(\cdot)||$ (solid line), and its relative change
$N/N_0 -1 \equiv ||u(\cdot,t)||/||f(\cdot)||-1$.  (dotted line).  Note that the
relative change in the norm is below one percent at $t = 200 \pi$,
that is 100 crossing times and $\approx 19,000$ iterations.}
\label{admfig}
\end{figure}

To test the algorithm, we give as initial data the following,
evaluated at  $t=0$
\begin{mathletters}
\label{admppsol}
\begin{eqnarray}
   \gamma_{ij} &=& 
              \left[A \left(m_{i} m_{j} - n_{i} n_{j} \right)
                  + B \left(m_{i} n_{j} + n_{i} m_{j} \right) \right]
               \sin (k\! \cdot\! x - \omega t) + \delta_{ij} 
\\
   K_{ij} &=& 
              \left[A \left(m_{i} m_{j} - n_{i} n_{j} \right)
                  + B \left(m_{i} n_{j} + n_{i} m_{j} \right) \right]
               \cos (k\! \cdot\! x - \omega t) \frac{\omega}{2} 
\end{eqnarray}
\end{mathletters}
where $m^{i}$ and $n^{i}$ are unit vectors orthogonal to the
propagation  vector $k^{i}$, i.e. $m\!\cdot\! m=1$, $n\!\cdot\! n=1$,
$m\!\cdot\! n=0$, $m\!\cdot\! k=0$ and $n\!\cdot\! k=0$. For
$k=(m1,m2,m3)$ with $m1,m2,m3$  integers, Eq.~(\ref{admppsol}) is a
periodic solution of the linear system  obtained by setting
$\gamma^{ij}=\delta^{ij}$ in~(\ref{admeqs}). During the integration,
we monitor the Hamiltonian constraint, given by
\begin{equation}
   H = R + K^2 - K_{ij} K^{ij}
\label{hamc}
\end{equation}
and, as a measure of the stability of the algorithm, the norm given
in Eq.~(\ref{admL2norm}). We have followed the evolution of a pulse
with $A=B=10^{-6}$, $k=(1,2,1)$, $\omega=\sqrt{6}$, taking
$m=(-7,4,1)/\sqrt{66}$ and $n=(-1,-1,3)\sqrt{6/11}$. The solution is
periodic in time as well, with period $T=2\pi/\sqrt{6}$, which
determines the natural time scale for the test problem. The equations
can be integrated on a grid of $(48)^3$ points for hundreds of
crossing times, without any signs of instability, as evident from
Fig.~\ref{admfig}, which shows $||u(\cdot,t)||/||f(\cdot)||$, 
the norm of the numerical solution as
a function of time, relative to the initial norm. The 
relative change in the norm,
with respect to the initial norm, 
is shown as well, and is below one percent
at 100 crossing times and $\approx 19,000$ iterations.

\begin{figure}
\centerline{\epsfysize=4in\epsfbox{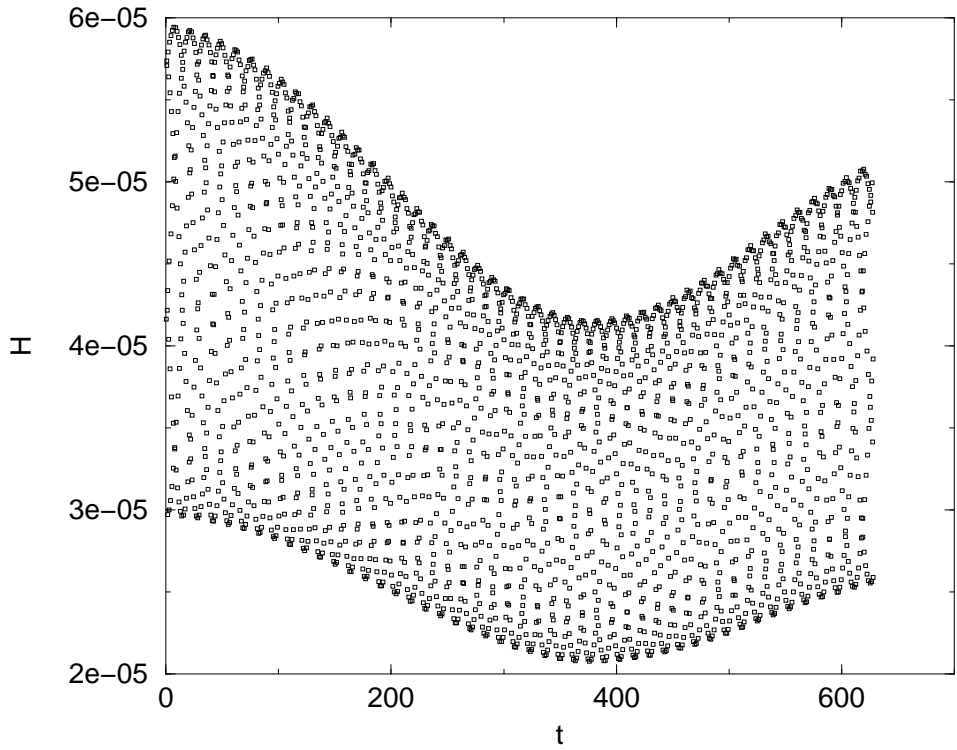}}
\caption{The maximum absolute value over the grid of the Hamiltonian 
constraint as a function of time for initial data corresponding to a 
linearized, periodic pulse.}
\label{hamcfig}
\end{figure}

We
have plotted in Fig.~\ref{hamcfig} the maximum absolute value over
the grid of the Hamiltonian constraint, Eq.~(\ref{hamc}) as a
function of time, and this quantity remains bounded throughout the
evolution, for upwards of 19,000 time steps. 

We have not carried out the Von Neumann stability analysis for the
principal part of the ADM equations, as we did for the simpler case
of the wave equation in subsec.~\ref{subsec:wavenum}, since it would be rather
involved and well beyond the scope of this work. To our knowledge,
this type of analysis has not been made for the other systems
mentioned here, either.

%---------------------------------------------------------------------
\section{Remarks} \label{sec:4}

%---------------------------------------------------------------------

In the first place, it is essential to emphasize that we have kept
ourselves within the context of problems with periodic boundaries, for
considerations of analytical and numerical stability alike.  In the case
of the Einstein equations, more often than not instabilities develop in
the course of a numerical simulation, but the origin of the instabilites
is not known, being attributed intermittently to either the equations
themselves or the particular boundary conditions being used. In the line
of a number of authors (most
recently,~\cite{admlong1,admlong2,admlong3,admlong4}) who have used the
standard ADM equations in exact (nonlinear) form and their
conformally-decomposed versions in numerical relativity, including
evolution in the strong-field regime for blackhole mergers, we have
presented a long-running stable code for the exact standard ADM
equations up to (un)stable boundaries.  With such a code, which runs for
sufficiently long times with periodic boundaries even if the equations
themselves are ill posed, the origin of instabilitites, if they occur,
can be shifted to the boundaries.  A discretization such as that used in
subsection~\ref{subsec:admnum} might be used to identify stable boundary
conditions other than periodic, or it might be used for matching to a
stable exterior characteristic code~\cite{characlong,highpower,cce}, in
which case the boundary values are provided by the exterior code and
are, in principle, consistent with the interior solution.  

We have not been involved with constraint propagation in this work,
but in this respect it is worth emphasizing that not only does the
code run for long times for the standard ADM equations, but it does
so preserving the hamiltonian constraint as well, in spite of the
ill-posedness of the evolution equations.     

Secondly, we have shown here that, from the analytical point of view, there
is no advantage to the conformally-decomposed equations~\cite{BS99} with
respect to the standard ADM form~\cite{york79}. Although a striking
difference between the numerical integration of the two systems has been
reported~\cite{BS99}, the origin of the difference in numerical behavior
must lie in some factor other than well-posedness (see~\cite{bspre} for a
discussion of other factors).  It might be thought that both systems may
have different properties when turned to full first-order form, and that
this might explain the difference in numerical behavior. So far this is an
open question. In this respect, it has been shown that reducing the
conformally-decomposed system (\ref{bs}) down to full first-order form by
defining the spatial derivatives  of the metric as new first-order
variables does not automatically make it well-posed, unless the lapse
function is densitized and the constraints are combined in specific ways
with the evolution equations~\cite{bspre}. Nevertheless, in reducing to
first-order form, there is plenty of freedom in the choice of first-order
variables. It is not clear at this time whether or not there exists a
choice of first-order variables for the reduction of (\ref{bs}) which will
turn the system into a well-posed one without densitizing the lapse or
combining with the constraints. The same might be said about the standard
ADM form, Eqs.~(\ref{admeqs}).

\acknowledgments

This research originated from stimulating conversations with Richard
Matzner with regards to unusual properties of the wave equation. We have
also benefited from conversations with Jeffrey Winicour and Mijan Huq. We
are grateful to Oscar Reula for his valuable impressions upon reading an
early draft of this work. This work has been supported by NSF under grants
No. PHY-9803301 to Duquesne University and No. PHY-9800731 to the
University of Pittsburgh.

%\bibliography{references}	% run latex, bibtex, latex, latex

\begin{thebibliography}{10}

\bibitem{york79}
J.~W. York,  in {\em Sources of Gravitational Radiation} (Cambridge University
  Press, Cambridge, 1979).

\bibitem{kurki-laguna-matzner93}
H. Kurki-Suonio, P. Laguna, and R.~A. Matzner, Phys. Rev. D {\bf 48},  3611
  (1993).

\bibitem{shibata}
M. Shibata and T. Nakamura, Phys. Rev. D {\bf 52},  5428  (1995).

\bibitem{BS99}
T.~W. Baumgarte and S.~L. Shapiro, Phys. Rev. D {\bf 59},  024007  (1999).

\bibitem{masso92}
C. Bona and J. Masso, Phys. Rev. Lett. {\bf 68},  1097  (1992).

\bibitem{newtonian}
S. Frittelli and O.~A. Reula, Commun. Math. Phys. {\bf 166},  221  (1994).

\bibitem{masso95}
C. Bona, J. Masso, E. Seidel, and J. Stela, Phys. Rev. Lett. {\bf 75},  600
  (1995).

\bibitem{york}
A. Abrahams, A. Anderson, Y. Choquet-Bruhat, and J.~W. York, Phys. Rev. Lett.
  {\bf 75},  3377  (1995).

\bibitem{moletter}
S. Frittelli and O.~A. Reula, Phys. Rev. Lett. {\bf 76},  4667  (1996).

\bibitem{helmut96}
H. Friedrich, Class. Quantum Grav. {\bf 13},  1451  (1996).

\bibitem{bspre}
S. Frittelli and O.~A. Reula, Well-posed forms for the conformally-decomposed
  3+1 {E}instein equations, to appear in J. Math. Phys. (October 1999).

\bibitem{fixing}
A. Anderson and J. James W.~York, Fixing Einstein's equations, preprint,
  gr-qc/9901021.

\bibitem{kreissbook}
B. Gustaffson, H.-O. Kreiss, and J. Oliger, {\em Time-dependent problems and
  difference methods} (Wiley, New York, 1995).

\bibitem{courant}
R. Courant and D. Hilbert, {\em Methods of Mathematical Physics} (Interscience
  Publishers, New York-London, 1962), Vol.~II.

\bibitem{hadamard}
A similar example is found in p. 33 of J. Hadamard, 
{\em Lectures on Cauchy's problem in linear partial differential
  equations} (Dover, New York, 1952).

\bibitem{recipes}
B.~P. Flannery, W.~H. Press, S.~A. Teukolski, and W.~T. Vetterling, {\em
  Numerical Recipes in Fortran 77}, 2nd  ed. (Cambride University Press,
  Cambridge, 1992).

\bibitem{admlong1}
G.~B. Cook {\it et~al.}, Phys. Rev. Lett. {\bf 80},  2512  (1998).

\bibitem{admlong2}
A.~M. Abrahams {\it et~al.}, Phys. Rev. Lett. {\bf 80},  1812  (1998).

\bibitem{admlong3}
B. Bruegmann, Int. J. Mod. Phys. D {\bf 8},  85  (1999).

\bibitem{admlong4}
M. Alcubierre {\it et~al.}, A conformal hyperbolic formulation of the
  {E}instein equations, preprint, gr-qc/9904013.

\bibitem{wald84}
R.~M. Wald, {\em General Relativity} (The University of Chicago Press, Chicago,
  1984).

\bibitem{characlong}
R. Gomez {\it et~al.}, Phys. Rev. Lett. {\bf 80},  3915  (1998).

\bibitem{highpower}
N.~T. Bishop {\it et~al.}, Phys. Rev. D {\bf 56},  6298  (1997).

\bibitem{cce}
N.~T. Bishop, R. G\'{o}mez, L. Lehner, and J. Winicour, Phys. Rev. D {\bf 54},
  6153  (1996).

\end{thebibliography}
%\bibliographystyle{prsty}

\newpage

\end{document}